\documentclass[conference]{IEEEtran}
\IEEEoverridecommandlockouts
\usepackage{cite}
\usepackage{amsmath,amssymb,amsfonts}
\usepackage{subfigure}
\usepackage{algorithmic}
\usepackage{graphicx}
\usepackage{textcomp}
\usepackage{xcolor}
\usepackage{soul}
\usepackage{hyperref}
\usepackage{booktabs}
\usepackage[utf8]{inputenc}
\usepackage{url}

\renewcommand{\v}[1]{\boldsymbol{#1}}
\newcommand{\argmin}{\operatornamewithlimits{argmin}}
\def\BibTeX{{\rm B\kern-.05em{\sc i\kern-.025em b}\kern-.08em
    T\kern-.1667em\lower.7ex\hbox{E}\kern-.125emX}}
\def\boxit#1{\vbox{\hrule\hbox{\vrule\kern6pt
          \vbox{\kern6pt#1\kern6pt}\kern6pt\vrule}\hrule}}

\begin{document}

\title{Rating Players of Counter-Strike: Global Offensive Based on Plus/Minus value\\
\thanks{}
}

\author{\IEEEauthorblockN{Hongyu Xu}
\IEEEauthorblockA{\textit{School of Mathematics and Statistics} \\
\textit{The University of New South Wales}\\
Sydney, Australia \\
z5415185@zmail.unsw.edu.au}
\and
\IEEEauthorblockN{Sarat Moka}
\IEEEauthorblockA{\textit{School of Mathematics and Statistics} \\
\textit{The University of New South Wales}\\
Sydney, Australia \\
s.moka@unsw.edu.au}
}

\maketitle
\begin{abstract}
We propose a player rating mechanism for Counter-Strike: Global Offensive (CS:GO), one of the most popular e-sport games, by analyzing players’ {\em Plus/Minus values}. The Plus/Minus value of a player is the average point difference between the player’s team and their opponent's over all the matches that the player involved.  By employing models including regularized linear regression, logistic regression and Bayesian linear models, we model the relationship between player participation and teams' point differences. 
Currently, most popular and influential metric for evaluating players in CS:GO community is ``Rating2.0". Unfortunately, this rating is based on players' individual scores only and doesn't capture player indirect contributions to their team's success. 
Our approaches provide a new rating mechanisms that effectively estimate direct and indirect contribution of each player towards the success of their team. 
In particular, our rating mechanisms prioritizes players who make a difference in winning matches rather than prioritizing the players who have higher individual scores. This would help clubs distribute prizes more fairly and recruit members in a more justifiable way. We believe that our methodology will have positive impacts on not only the CS:GO community but also the whole e-sports industry. 
\end{abstract}


\section{Introduction}
E-sports, short for electronic sports, has experienced rapid growth over the years despite facing many skepticism. The first-person shooter game {\em Counter-Strike} released in 2000 marked a pivotal moment for e-sports being popular all around the world. The World Cyber Games (WCG) established in 2000 is an event organized like the Olymics. Counter-Strike entered WCG in 2001 and became an important member of WCG \cite{hutchins2008signs}. Despite setbacks such as the discontinuation of WCG in 2014 due to economic crisis, e-sports continued its trajectory \cite{julian2017esports}. An example highlighting the growth of e-sports is The International Dota 2 Tournaments in 2021 with a prize pool exceeding 40 million US dollar \cite{ti102021} rivaling prestigious sporting events like Wimbledons total prize money of £35 million that same year \cite{wimbledon}. By 2023 several electronic games had become medal events in the Hangzhou Asian Games.

A surprising fact is that the viewership of e-sports events surpassing one billion viewers, which far exceeds the audience size of traditional sports events in terms of views. However, e-sports clubs face challenges in generating revenue from ticket sales and merchandise at venues since most e-sports matches are held in rented venues than club owned facilities. According to {\em PricewaterhouseCoopers} (PwC), on average an e-sports fan spends around \$3.60 on e-sports compared to \$54 spent by sports enthusiasts. E-sports heavily rely on sponsorship income, which makes up a portion of team revenues ranging from 40 to 90 percent \cite{mangeloja2019economics}. Consequently, the e-sports clubs may not be able to operate even if only a few sponsors stop supporting it, negatively impacting the stability and value of e-sports competitions.

One key reason of deficit of e-sports clubs is the lack of data analytics in the e-sports industry comparing to traditional sports \cite{rubleske2020sports}. Although some studies have tried to evaluate teams \cite{pradhan2020power} or individual players \cite{makarov2018predicting}, commercial applications of these approaches are still limited. As suggested by \cite{Hvattum2019ACR}, implementing accurate evaluation methods for players could enhance financial stability for e-sports clubs and boost market growth. 

Plus/Minus value of a player is the average point difference between his team and the opponent team, where the average is computed over all the matches that the player played.
Contrary to traditional box scores (recordings of each player's individual achievements), Plus/Minus value will not be impacted by the judgment of the scorekeeper, which can introduce subjective biases and inconsistencies~\cite{van2017adjusting}. Furthermore, it can comprehensively capture a players performance on the court \cite{sisneros2013expanding}. The evaluation of players based on their Plus/Minus value scores has been widely used in many traditional sports such as basketball \cite{sill2010improved}, football \cite{schultze2018weighted} and hockey \cite{macdonald2011regression}. Its effectiveness has been proven in these traditional sports but has not yet been commonly applied in analyzing e-sports data.

\begin{figure}[h]
    \centering
    \includegraphics[width=\linewidth]{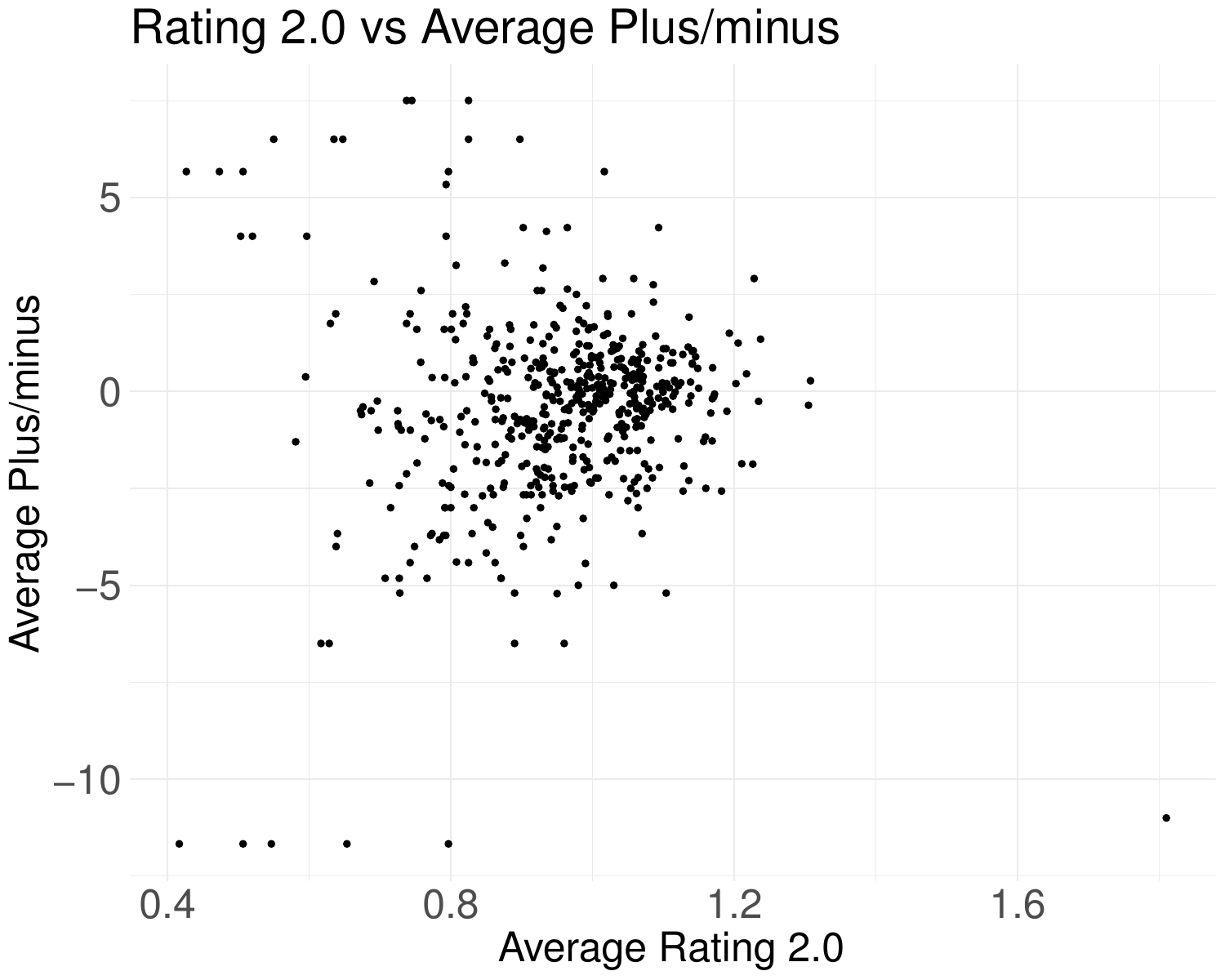}
    \caption{Relationship between players' Average Rating2.0 and player's  Plus/Minus value.}
    \label{fig:rating20}
\end{figure}

Counter-Strike: Global Offensive (CS:GO) is the sequel to the renowned Counter-Strike and it is the most popular first-person shooter game in the world of e-sports. For the CS:GO games, {\em Rating2.0} is the popular metric currently used for evaluating players and it significantly influences how players are perceived in CS:GO community. 
Top accolades often given to individuals with top Rating2.0 scores. This metric also occupy the dominant position when doing team recruitment decisions. However, Rating2.0 is a scoring metric based on box score which have many limitations. Based on the data obtained from {\em Half-Life Television} (HLTV) covering all big events matches from 2018 to 2023,  the scatter plot in Figure \ref{fig:rating20} illustrates the connection between players’ Average Rating2.0 and their Plus/Minus value. A hypothesis test using the Pearson correlation coefficient is conducted to investigate the relationship between these factors. The $p$-value of the hypothesis test is $0.293$ which indicates that we cannot reject the null-hypothesis suggesting no significant association between a player’s Rating2.0 and his Plus/Minus value. This analysis illustrates the lack of correlation between a player contribution to his team in rounds during games and their Rating2.0 value. Many ways that a player can contribute to their teams’ success is neglected if we use only Rating2.0 to assess a player.

Our study aims to develop a rating mechanism based on Plus/Minus values to assess a player's overall impact rather than just their box scores. 
Our first step is to investigate how players contribute to team victories by using a linear model inspired by Rosenbaum \cite{APM}, who analyzed how a player’s appearance in the game impacts the teams point difference. Afterwards, we provide regularized linear models, logistic regression models and Bayesian linear models to overcome certain difficulties in e-sports. These difficulties include variable selection and multicollinearity in player appearance. We trained our models using the data of big event matches from 2018 to 2022. The ratings of players from Bayesian model and elastic net logistic model are not only correlated to the Plus/Minus value from 2018 to 2022 but also correlated to the Plus/Minus value in 2023. This indicates that the rating mechanism based on Plus/Minus value captures each player's contribution to their team while exhibiting the power of predicting players' future performances. We believe that this new approach could enhance award selections and salary evaluations in the CS:GO industry, ultimately benefiting the growth of CS:GO and e-sports industry. 

The remaining paper is organized as follows: In Section~\ref{sec:data}, we  explain the
source and structure of our data. In Section~\ref{sec:proposed-methods}, we introduce our models based on Plus/Minus values. In Section~\ref{sec:results}, we discuss the results of these models and compare them with Rating2.0. Finally, in Section~\ref{sec:conclusion}, we conclude the paper and discuss some future works.

\section{Data}
\label{sec:data}
\subsection{Introduction of CS:GO Game}
\label{subsec:CSintro}
Matches of CS:GO consist of 30 rounds with victory going to the team that wins 16 rounds first. In case of a tie when both teams have 15 wins, some matches allow for a draw while most proceed to an overtime period comprising 6 rounds. The first team to secure 4 rounds in overtime will win the match. If both teams win 3 rounds, there will be another overtime until one team attains a total of 4 wins in one overtime period. Each team comprises five players and for each round the attacking team have the opportunity to secure victory either by planting a bomb or by eliminating all members in defending team. Conversely the defending team can achieve success by defusing the bomb or by eliminating all members in attacking team. Following 15 rounds of play, the teams switch roles.

\subsection{The Source of Our Data}
\label{subsec:DTsource}
There are many CS:GO competitions held each year offering a championship prize pool exceeding \$500,000. Professional teams in CS:GO aim to get medals in these events not only to 
win the prize money but also to garner fan support by increasing their visibility. The data we used in this research is sourced from the {\em hltv.org}, a platform that has kept demos and match details of CS:GO tournaments since 2012. Our dataset encompasses all matches that was classified as big events by HLTV spanning from 2018 to 2023 including teams’ scores, players’ appearance and players’ box scores. \footnote{For more details about the data, please visit the GitHub repository at \url{https://github.com/patchoulixu/hltv-data}.}

\begin{figure*}[htbp]
    \centering
    \subfigure[Distribution of {\em ``\texttt{ResultDiff}".}]{
        \includegraphics[width=0.22\linewidth]{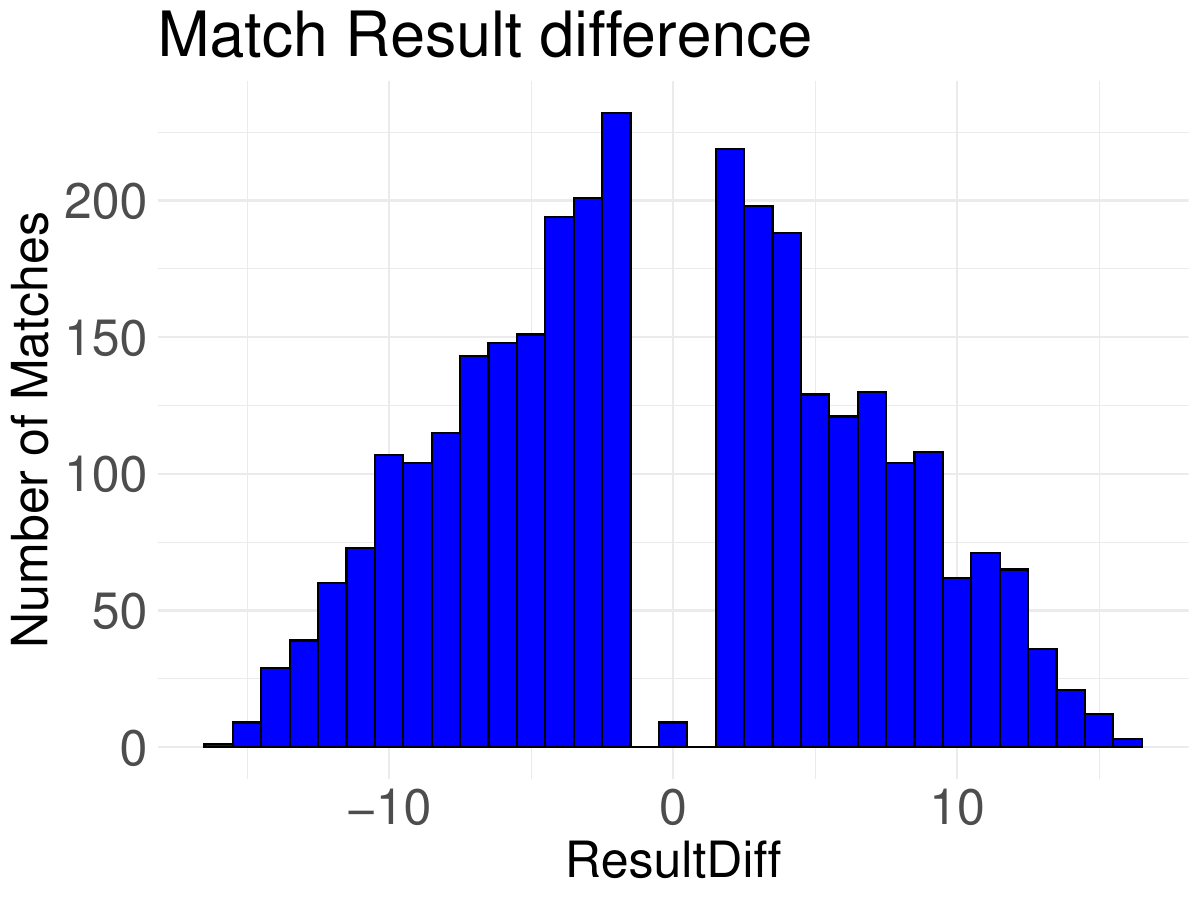}
        \label{fig:resultdiff}
    }
    \subfigure[Yearly average of absolute value of {\em ``ResultDiff"} with its variability given by $95\%$ confidence intervals.]{
        \includegraphics[width=0.22\linewidth]{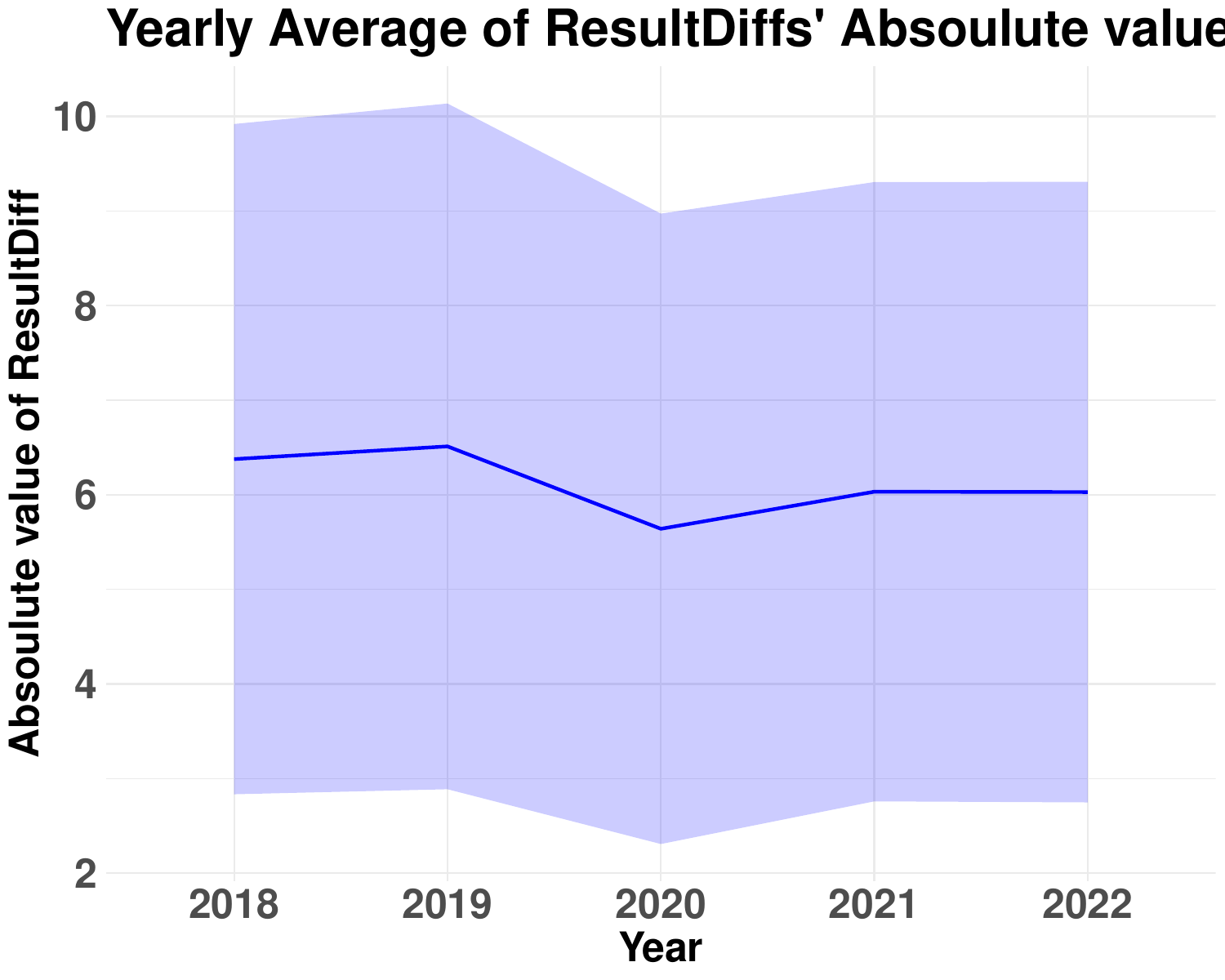}
        \label{fig:yeargraph}
    }
    \subfigure[Distribution of average Rating2.0 and Plus/Minus value of players. ]{
        \includegraphics[width=0.22\linewidth]{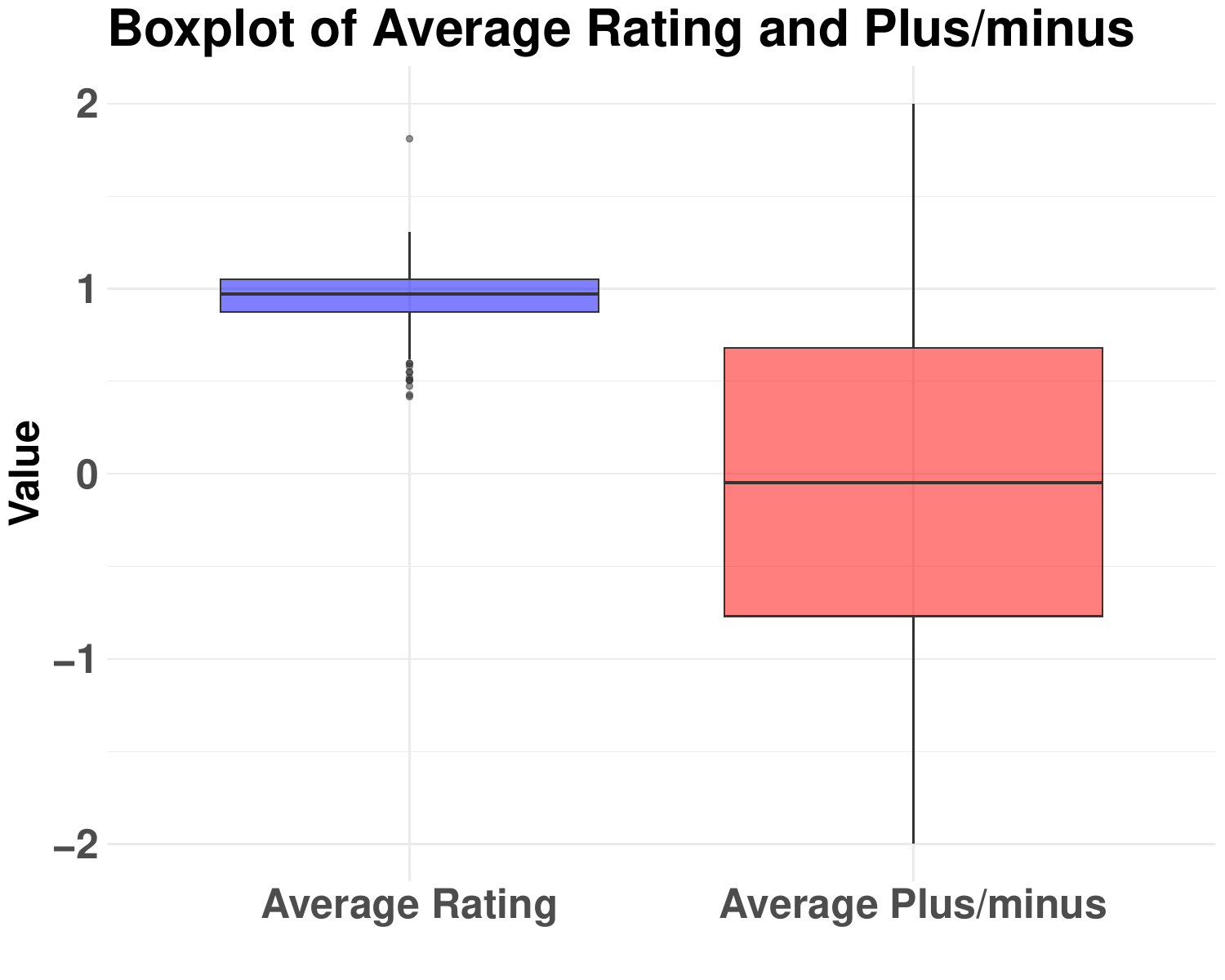}
        \label{fig:boxpm}
    }
    \subfigure[Scatterplot for comparing the Plus/Minus value and the number of matches attended.]{
        \includegraphics[width=0.22\linewidth]{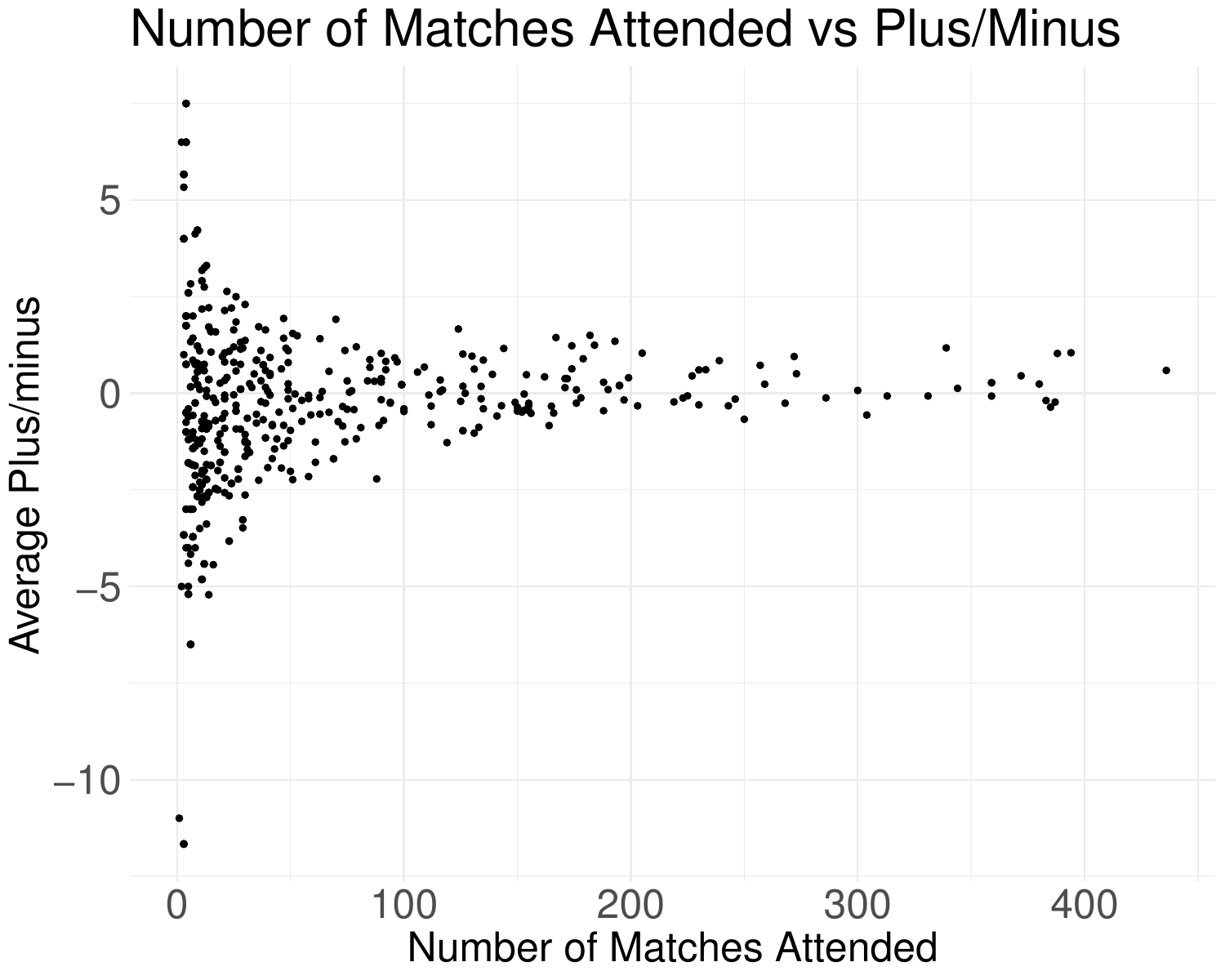}
        \label{fig:attendpm}
    }
    \caption{Some features of player performance data.}
    \label{fig:allfigs}
\end{figure*}

\subsection{The Structure of the Data}
\label{subsec:DTintro}

We have organized the data into two sets to help us develop the regression models. The first set of data records each CS:GO match as a row, containing details on player involvement and scores. Within this set, ``\texttt{MapId}" is the primary key of our data frame. The ``\texttt{ResultDiff}" variable captures the point difference (Team 1's score minus Team 2’s score) for every match, acts as the response vector in our model. Furthermore the appearance data of over 500 players was recorded in this dataset and serves as the design matrix for our model. If a player’s appearance value is $1$ it means he is playing for Team~1. A value of $0$ indicates the player did not participate in that match while~$-1$ signifies the player playing for Team~2. 

To compute the Plus/Minus value of {\em Player}-$j$, for the $i$-th match played by {\em Player}-$j$, let $r_i$ be the \texttt{ResultDiff} of that match and 
\[
s_{j, i} 
= 
\begin{cases}
r_i, &\quad \text{if {\em Player}-$j$ is in Team 1},    \\
-r_i, &\quad \text{if {\em Player}-$j$ is in Team 2}.
\end{cases}
\]
Then, the Plus/Minus value is defined as the average of $s_{j, i}$'s over all the matches played by {\em Player}-$j$.

The second dataset includes the values of Rating2.0 for players in each game. These scores were gathered to compute the average Rating2.0 which can be compared with our ratings or used as a prior information in Bayesian methods. In Table~\ref{tab1}, we show how the response vector and design matrix are structured.

\begin{table}[htbp]
\caption{Example of the Response Vector and Design Matrix}
\begin{center}
\begin{tabular}{|c|c|c|c|c|c|}
\hline
\textbf{MapID} & \multicolumn{5}{|c|}{\textbf{Data of match}} \\
\cline{2-6} 
 & \textbf{\textit{ResultDiff}}& \textbf{\textit{Player-1}}& \textbf{\textit{Player-2}}& \textbf{\textit{...}}& \textbf{\textit{Player-518}} \\
\hline
76059 & 4 & 1 & 0 & ... & -1 \\
76060 & -5 & -1 & 0 & ... & 1 \\
76061 & 7 & 1 & 0 & ... & -1 \\
... & ... & ... & ... & ... & ... \\
\hline
\end{tabular}
\label{tab1}
\end{center}
\end{table}

\subsection{The Features of the Data}
\label{subsec:DTFeature}

In Figure \ref{fig:resultdiff}, we display the distribution of point differences of single matches from 2018 to 2023. In CS:GO matches, a game ends only when one team wins by least two points more than the other, and as a result, point difference discrepancies cannot be $1$ or $-1$. Most games go into overtime until a victory is determined. However, a small number of games end-up in ties as shown in the figure with a point difference of $0$.
As depicted in the visuals apart from $1$, $0$ and $-1$, the distribution of point differences seems to follow a normal distribution. In Figure~\ref{fig:yeargraph}, it is clear that the mean and variability of the ``\texttt{ResultDiff}" values across the years remain fairly consistent. 

In the Figure \ref{fig:boxpm}, we notice that the average Rating2.0 for players is close to 1 and has less variability, and in contrast, the Plus/Minus value scores for players is around~$0$ and has more variability. Figure \ref{fig:attendpm} compares each player's Plus/Minus value score and the number of matches they played. We observe that the mean of their Plus/Minus value scores does not change as players attend more matches, while the variability is decreasing. Players who attended only a few matches could be considered as outliers and need to take into account when constructing a regression model.

\section{Proposed Methods}
\label{sec:proposed-methods}

In this section, we present our methods to estimate each player contribution towards the success/failure of a match. These models consist of linear models, hierarchical Bayesian linear models, and logistic regression models. 


\subsection{Linear Regression Models}
\label{subsec:linear}
\subsubsection*{Motivation}

There are two main reasons for using linear regression models to analyse the the relationship between the Plus/Minus values and the performance of the players. First, multivariate linear regression is a model that aims to explore the relationships between multiple independent and dependent variables, allowing us to consider the influence of teammates and their opponents on the Plus/Minus values \cite{zhang2019multiple}. The second reason is that the linear models are relatively more interpretable than many other models that can explore the relationships between multiple independent and dependent variables. 
For instance, many modern models, such as neural networks, act like a black box \cite{weisberg2005applied}, making it difficult to explain the connection between player appearances and teams' point differences through such models, making it difficult to measure a player's contribution towards the success/failure of the match of using these models.

While providing high degree of interpretability, unfortunately, however, the multiple linear models without any regularization exhibit the issue of multicollinearity. In particular, many players often appear on the court simultaneously, and this can cause a high degree of correlation between the features in the data correspond to these players, which violates the assumptions required by the least squares method \cite{farrar1967multicollinearity}. To overcome this difficulty, we consider the {\em ridge regression} method, which is a commonly used approach to address the multicollinearity issue in multiple linear regression \cite{mcdonald2009ridge} by incorporating an $\mathcal{L}_2$ penalty in optimization objective. 


Another approach is to omit some columns in the data that are highly correlated with others \cite{daoud2017multicollinearity}. One popular approach for such feature selection is called {\em lasso regression}, which incorporate an $\mathcal{L}_1$ penalty in optimization objective \cite{tibshirani1996regression}. However, the lasso regression often doesn't perform as well as ridge regression when handling multicollinear data \cite{zou2005regularization}. {\em Elastic net} method, introduced by \cite{zou2005regularization}, is a generalization of both lasso and ridge regression, and it helps in handling multicollinear data while allowing feature selection. In this paper, we consider both ridge and elastic net regressions to mitigate the multicollinearity problems in the data.


\subsubsection*{Formulation of Multiple Linear Model}

A multiple linear model can be represented as 
\begin{equation}
\v{y} = X\v{\beta} + \v{\epsilon},
\end{equation}
where, $\v{y} = (y_1, \dots, y_n)$ is the the response vector with $y_i$ denoting the ``\texttt{ResultDiff}"  for the $i$-th game. That is, $y_i$ is the number of rounds won by one team (called Team~1) minus the number of rounds won by the other team (called Team~2) in $i$-th game.  The design matrix $X$ records the players' attendance. In particular, each column of $X$ corresponds to a player and each row corresponds to a game. The $(i, j)$-th element $x_{ij}=1$ indicates that player~$j$ played for Team~1 in $i$-th game, $x_{ij}=-1$ indicates that player~$j$ played for Team~2, and $x_{ij}=0$ indicates that the player $j$ did not play in the $i$-th game.
The coefficient vector $\v{\beta} = (\beta_1, \dots, \beta_p)$, where $p$ is the number of players and the $j$-th coefficient $\beta_j$ represents the effect of player $j$'s presence on the point difference between the two teams comparing to average players. As is evident, this model has no intercept term because without any contribution from players, the score must be zero. Finally, $\v{\epsilon}$ is assumed to be an additive mean-zero Gaussian noise vector with independent and identically distributed elements of unknown variance.

The goal in linear regression is to estimate the coefficient vector $\v \beta$ by solving the convex optimization problem given by
\[
\hat{\v \beta} =  \argmin_{\v \beta} \frac{1}{2n}\|\v y - X\v \beta \|_2^2,
\]
where $\| \cdot \|_2$ denotes the usual Euclidean norm.  The above problem has a solution given by 
\[
\hat{\v \beta} = \left( X^\top X\right)^\dagger (X^\top \v y),
\]
where $A^\dagger$ denotes the pseudo (i.e., Moore–Penrose) inverse of a matrix $A$. 

The elements of solution $\hat{\v \beta}$ provide ratings to the players and we rank the players according to the descending order of these ratings. Specifically, let $i_1, i_2, \dots, i_p$ are such that 
\[
\hat \beta_{i_1} \geq \hat \beta_{i_2} \geq \cdots \geq \hat \beta_{i_p},
\]
then the $i_1$-th player has rank 1, the $i_2$-th player has rank 2, so on. When two players have the same ratings, we break ties at random. 


\subsubsection*{Formulation of Regularized Linear Models}
We now formulate the elastic net regression. As mentioned earlier, this model is useful for handling data with multicolinearity while incorporating some sparsity in the solution. In addition, we obtain the ridge regression as a special case of the elastic net regression. 


The elastic net regression model can be stated as 
\begin{equation}
\hat{\v \beta} = \argmin_{\v \beta} \left(\frac{1}{2n} \| \v y - X\v \beta \|_2^2 +  \alpha \lambda \| \v \beta \|_1 + (1 - \alpha )\lambda \| \v \beta \|_2^2 \right),
\label{eqn:elastic-reg}
\end{equation}
where the regularization parameter $\lambda$ controls the strength of the regularization effect and the parameter $\alpha$ controls the mixing proportion of $\mathcal{L}_1$ and $\mathcal{L}_2$ penalty terms. We use the cross-validation approach to determine the values of the optimal penalty coefficient $\lambda$ and $\alpha$; for more details, refer to Section~\ref{sec:results}.

In particular, when $\alpha = 0$, the elastic regression in \eqref{eqn:elastic-reg} becomes ridge regression. That is, the solution of the ridge regression is given by 
\begin{equation*}
\hat{\v \beta} = \argmin_{\v \beta} \left(\frac{1}{2n} \| \v y - X\v \beta \|_2^2 + \lambda \| \v \beta \|_2^2 \right),
\end{equation*}
with $\lambda$ being the penalty parameter.

As detailed earlier, we rank the players based on the estimated coefficients (or, ratings) $\hat {\v \beta}$.

\subsection{Hierarchical Bayesian Model}
\label{subsec:bayes}
\subsubsection*{Motivation}
The primary goal of the {\em Bayesian models} is to incorporate prior knowledge about a problem into data analysis~\cite{gelman1995bayesian}. To distinguish between players that are highly correlated, we explore the Rating2.0 scores as a prior knowledge and this allows us to use Bayesian linear models. Since Rating2.0 scores are derived from box score metrics, 
effectiveness of Rating2.0 scores vary based on the players' positions in the game. 
To incorporate this variability in the Rating2.0 scores, we utilize hierarchical Bayesian models that adjust the prior distributions for each player using some hyperparameters. 
We apply the Markov Chain Monte Carlo (MCMC) technique to fit the Bayesian model \cite{gilks1995markov}.

\subsubsection*{Problem Setting}
First we consider a simple Bayesian multivariate linear model applying the Rating2.0 scores as the mean of prior distribution on the coefficients as follows:
\[
\v{y} \mid \v{\beta} \sim \mathcal{N}(X\v{\beta}, \, \sigma^2 I),
\]
with the coefficients $\v \beta$ assumed to have a normal distribution given by
\[
\v{\beta} \sim \mathcal{N}(\textsf{sRating2.0},\, \tau^2 I),
\]
for a fixed variance $\tau^2$, where {\sf sRating2.0} is the vector of standardized Rating2.0 scores of all the players.

We now turn to a hierarchical Bayesian model varying prior distributions among different player types. 
For example, some players take on a commanding role within the team, and their tactical leadership greatly influences the outcome of the game. These individuals prioritize directing team members over focusing on their performance. Given this, it wouldn't make sense to use the same prior distribution with the mean of Rating2.0 for these leaders who might have a low Rating2.0 score. As a remedy, we use hierarchical models which could recognizing both distinctions and similarities between player groups by multiple levels in the model.

The mathematical expression of the hierarchical model is as follows.
Given the design matrix $X$ and the response variable $\v y$,  we assume that 
\begin{align*}
\v{y} \mid \v{\beta} &\sim \mathcal{N}(X\v{\beta},\, \sigma^2 I),\\
\v{\beta} \mid \v{\eta} &\sim \mathcal{N}(\v{\eta} \odot \textsf{sRating2.0},\, \tau^2 I),\,\,\, \text{and}\\
\v{\eta} &\sim \mathcal{N}(\v 0,\, I),
\end{align*}
with $\odot$ denoting the Hadamard (i.e., element-wise) product between vectors, 
where $\v{\beta}$ is the rating vector, $\v{\eta}$ represents the differences in the impact of the \textsf{sRating2.0} vector on the prior distribution for different players.
We assumed that distribution of the hyperparameter vector~$\v{\eta}$ is a standard normal. 

In such a model, the mean $\v \eta \odot \textsf{sRating2.0}$ in the distribution of $\v{\beta} \mid \v{\eta}$ allows us to differentiate the impact on the point differences between player combinations that often appear together. Furthermore, through the hyperparameters of $\v \eta$, we consider the differences in the impact of Rating2.0 scores on the prior distribution for players of different positions.

\subsection{Logistic Regression}
\label{subsec:logistic}
\subsubsection*{Motivation}

As we mentioned, the final goal of competitive sports is to win the match. Recall that positive point difference (i.e., positive response value) means winning the match while negative point difference means losing. 
Given the concern about overfitting due to the complexity of utilizing point difference as the response variable, we aim to simplify the data by transforming the point difference into a win/loss binary scenario. Logistic regression has been widely employed 
since 1970 for binary response cases \cite{cramer2002origins}. Additionally, we considered logistic regression  with ridge and elastic net penalties to mitigate multicollinearity concerns in the design matrix \cite{pereira2016logistic, algamal2015regularized}.

\subsubsection*{Problem Setting}

In this section, we present three versions of logistic regression model: one without any penalty, one with ridge penalty (i.e., $\mathcal{L}_2$ regularization) and one with elestic net regularization.

We use the logit function to replace the point difference values to winning probabilities $\hat{\v p} = (\hat{p}_1, \dots, \hat{p}_n)$, 
where $\hat{p}_i$ is the winning probability of Team 1 in the $i$-th match given by 
\begin{equation}
\hat{\v p} = \frac{1}{1 + e^{-\v x_i^{\top} \v \beta}},
\end{equation}
where \(\v x_i\) is the $i$-th row of the design matrix $X$ and \(\v \beta\) is the coefficient vector, which we estimate by solving the convex optimization problem given by
\begin{equation}
\argmin_{\v \beta} \left\{ -\frac{1}{2n} \sum_{i = 1}^n y_i \log\hat{p}_i + (1 - y_i) \log(1 - \hat{p}_i) \right\},
\end{equation}
where \(n\) is the number of samples and \(y_i \in \{0, 1\}\) with~$1$ indicating the winning of Team~1 in the $i$-th match and~$0$ indicating the loss. As mentioned earlier, when we apply logistic regression, we ignore the matches that resulted in draw. 

Similar to the linear regression, the ridge logistic regression is a special case of the elastic net logistic regression. In particular, the elastic net logistic regression model can be stated as
\begin{equation}
\begin{split}
\argmin_{\v \beta} \Bigg\{ -\frac{1}{2n} \sum_{i = 1}^n y_i \log\hat{p}_i + (1 - y_i) \log(1 - \hat{p}_i)  \\
 + \alpha \lambda \|\v \beta\|_1 + (1-\alpha)\lambda \|\v \beta\|_2^2 \Bigg\}
\end{split}
\label{eqn:elstic-logistic}
\end{equation}
where $\alpha$ is the parameter that controls the mixing proportion of $\mathcal{L}_1$ and $\mathcal{L}_2$ regularization terms and \(\lambda\) is  the regularization strength. In the specific case, when $\alpha = 1$, \eqref{eqn:elstic-logistic} becomes the ridge logistic regression.

\section{Results}
\label{sec:results}
In this section, we first discuss the parameters we choose in our simulations. Then,  we compare the performance of the different models considered earlier using scatter plots of the predicted Plus/Minus value and the true Plus/Minus value of the players. 

\subsection{Model Parameter Configuration}
In this part, we discuss the choice of parameters used in our simulations. 
The first parameter is the minimum number of games that a player must participate in to be part of the data. As shown in \ref{fig:attendpm}, the variability of player's Plus/Minus value is too large when they only attended a few matches. Thus, we exclude players who play less than 50 matches from the data to avoid the problem of overfitting.

In regularized linear regression models and logistic regression models, we use 10-fold cross-validation approach to determine the values of the optimal penalty coefficients $\lambda$ and~$\alpha$ on a grid. To construct the grid, we selected $100$ equally placed values on $[0,1]$ for $\alpha$ and the default grid provided by the \texttt{cv.glmnet} function from the R package \texttt{glmnet} for~$\lambda$. We randomly split the data to have $80\%$ for training and $20\%$ for testing.

We utilized the \texttt{Stan} programming language to fit the Bayesian model.  
\footnote{\texttt{Stan} is a powerful probabilistic programming language designed for statistical inference, and it is implemented in C++. More information can be found on the official website of Stan: \url{https://mc-stan.org/}.}

\subsection{Comparing Proposed Ratings with Rating2.0}
As mentioned earlier, we believe that a key factor in assessing a player's performance is their impact on the teams’ point differences (equivalently, on winning and losing). Hence, we believe that a good rating metric should correlate to the teams’ point differences when the player is playing for the team.

By using our new ratings derived from the regression coefficients of the proposed models, we estimate Plus/Minus value of players. If there is a significant correlation between a player's estimated Plus/Minus value and their true Plus/Minus value (in logistic model it is correlation between estimated win rate and actual win rate of players), we conclude that our rating methods are more accurate in reflecting players contributions to their teams performance than Rating2.0.

\begin{figure*}[!htbp]
    \centering
    \subfigure[]{
        \includegraphics[width=0.22\linewidth,height=0.15\textheight]{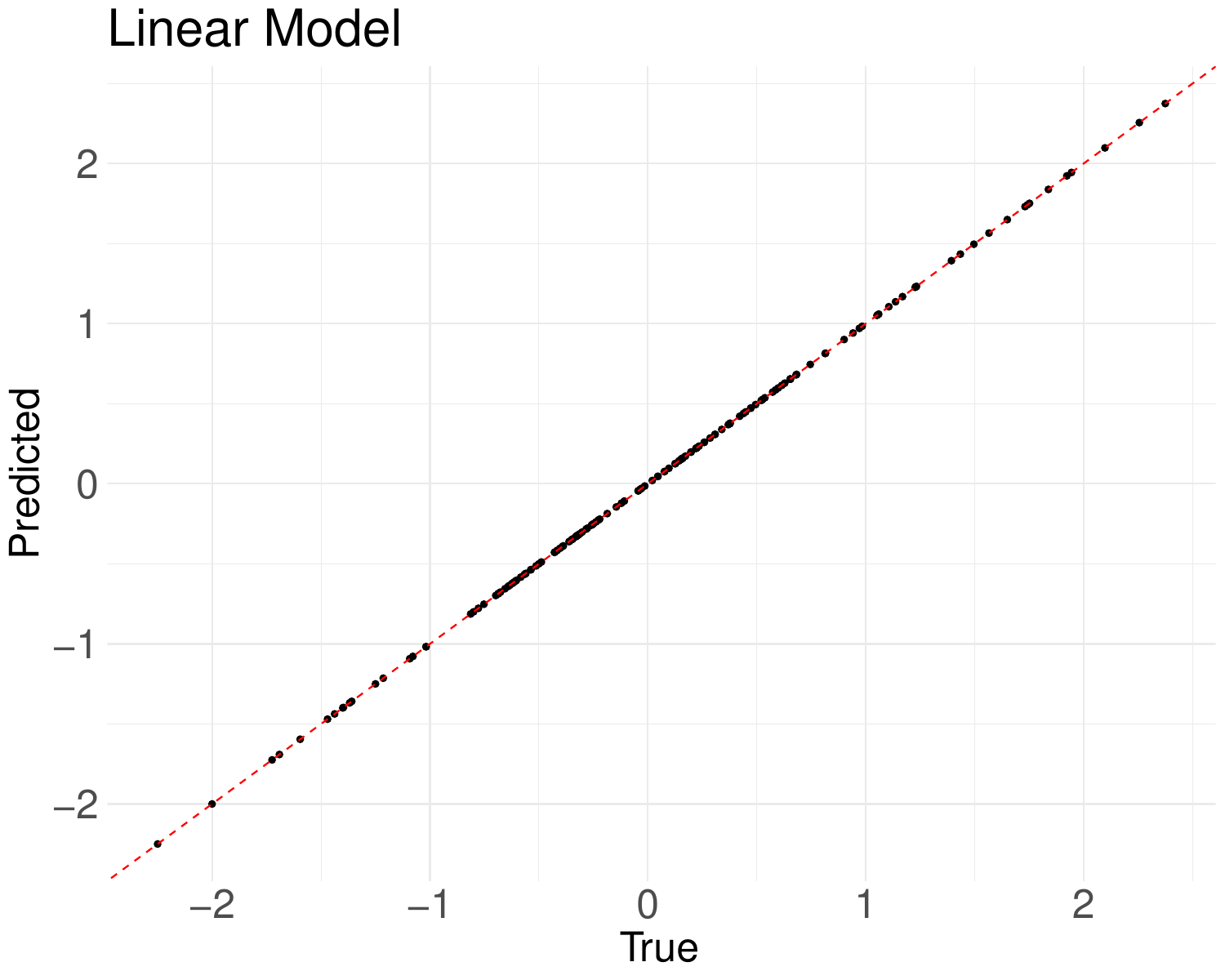}
        \label{fig:lineartrain}
    }
    \hfill
    \subfigure[]{
        \includegraphics[width=0.22\linewidth,height=0.15\textheight]{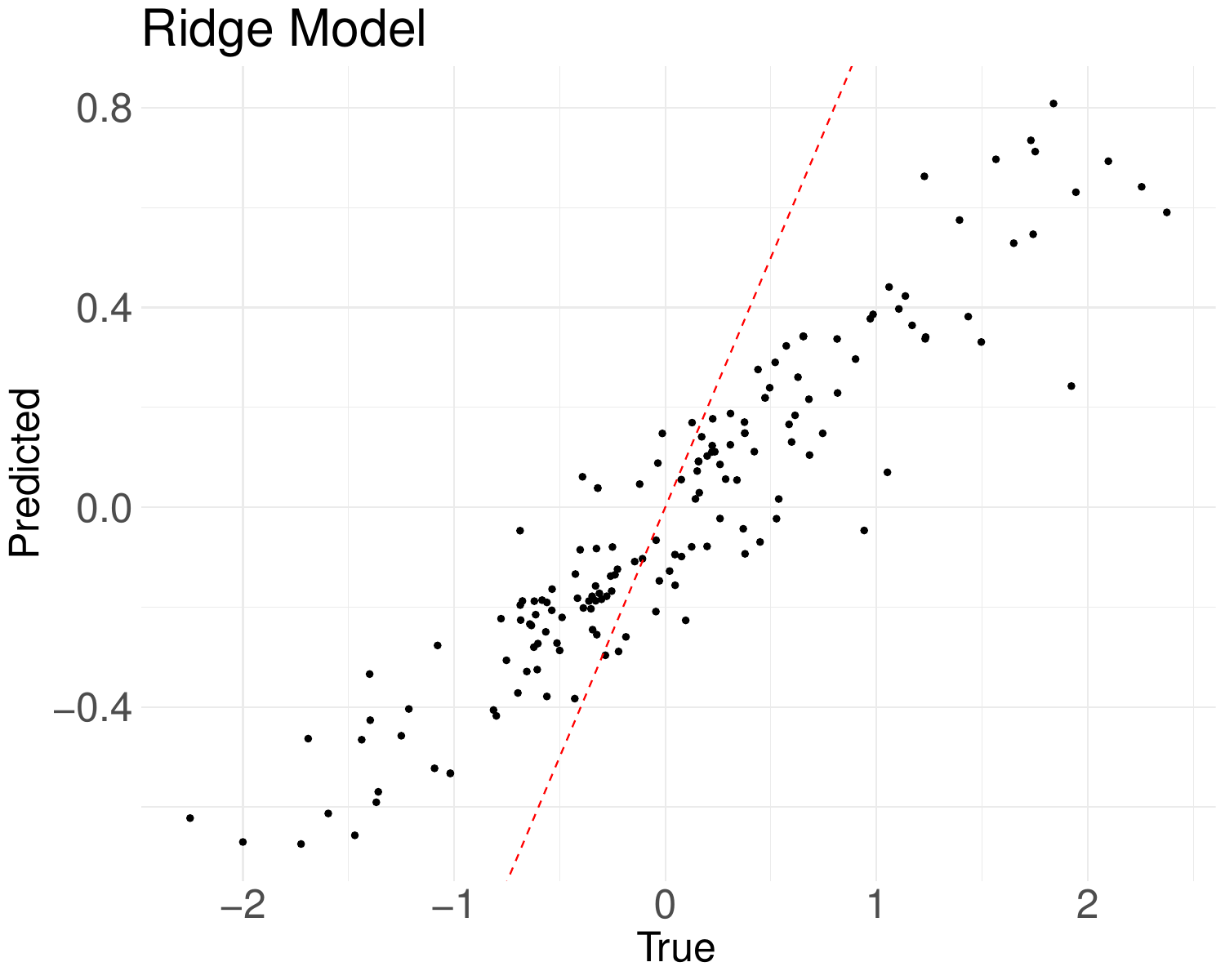}
        \label{fig:ridgetrain}
    }
    \hfill
    \subfigure[]{
        \includegraphics[width=0.22\linewidth,height=0.15\textheight]{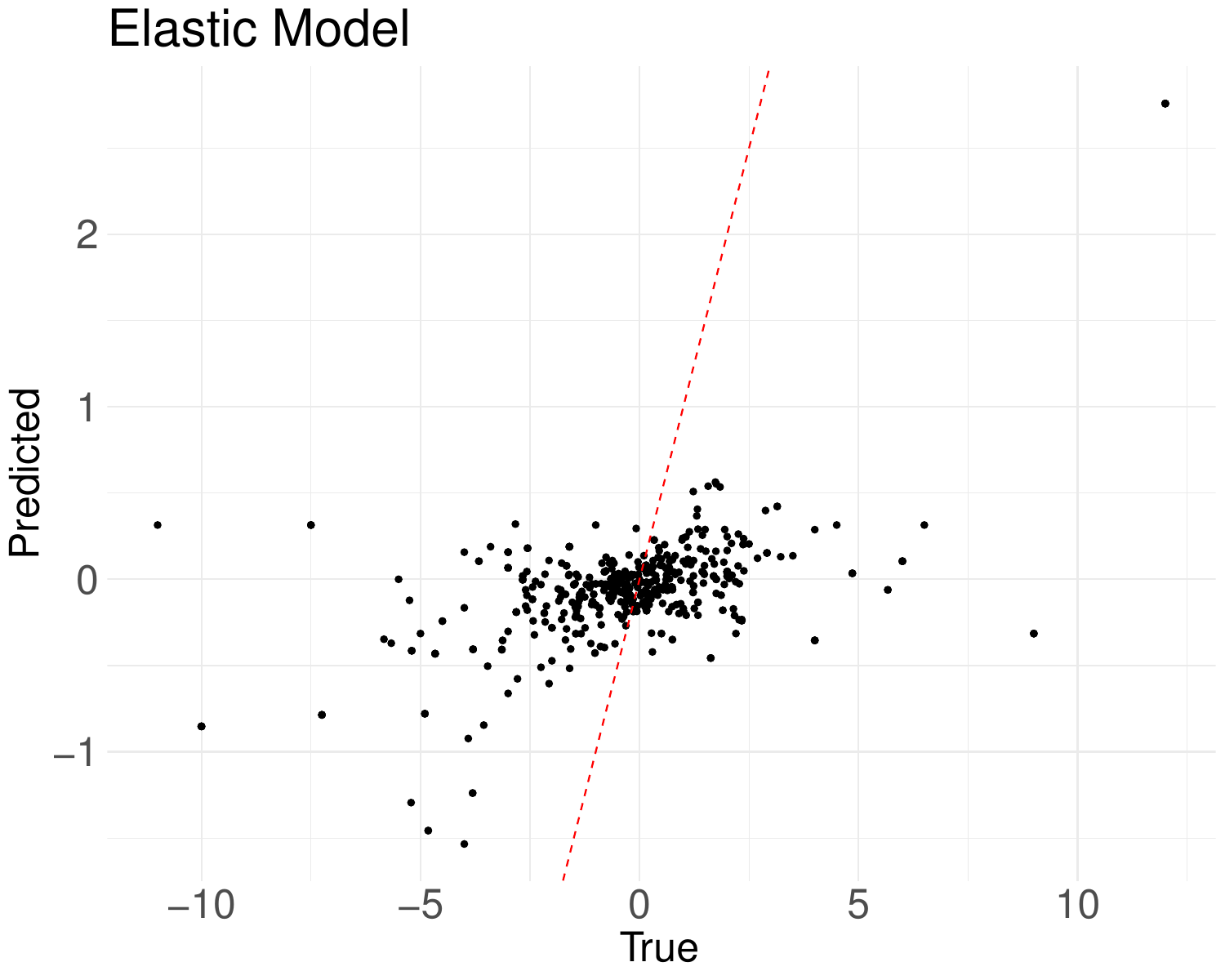}
        \label{fig:elatrain}
    }
    \hfill
    \subfigure[]{
        \includegraphics[width=0.22\linewidth,height=0.15\textheight]{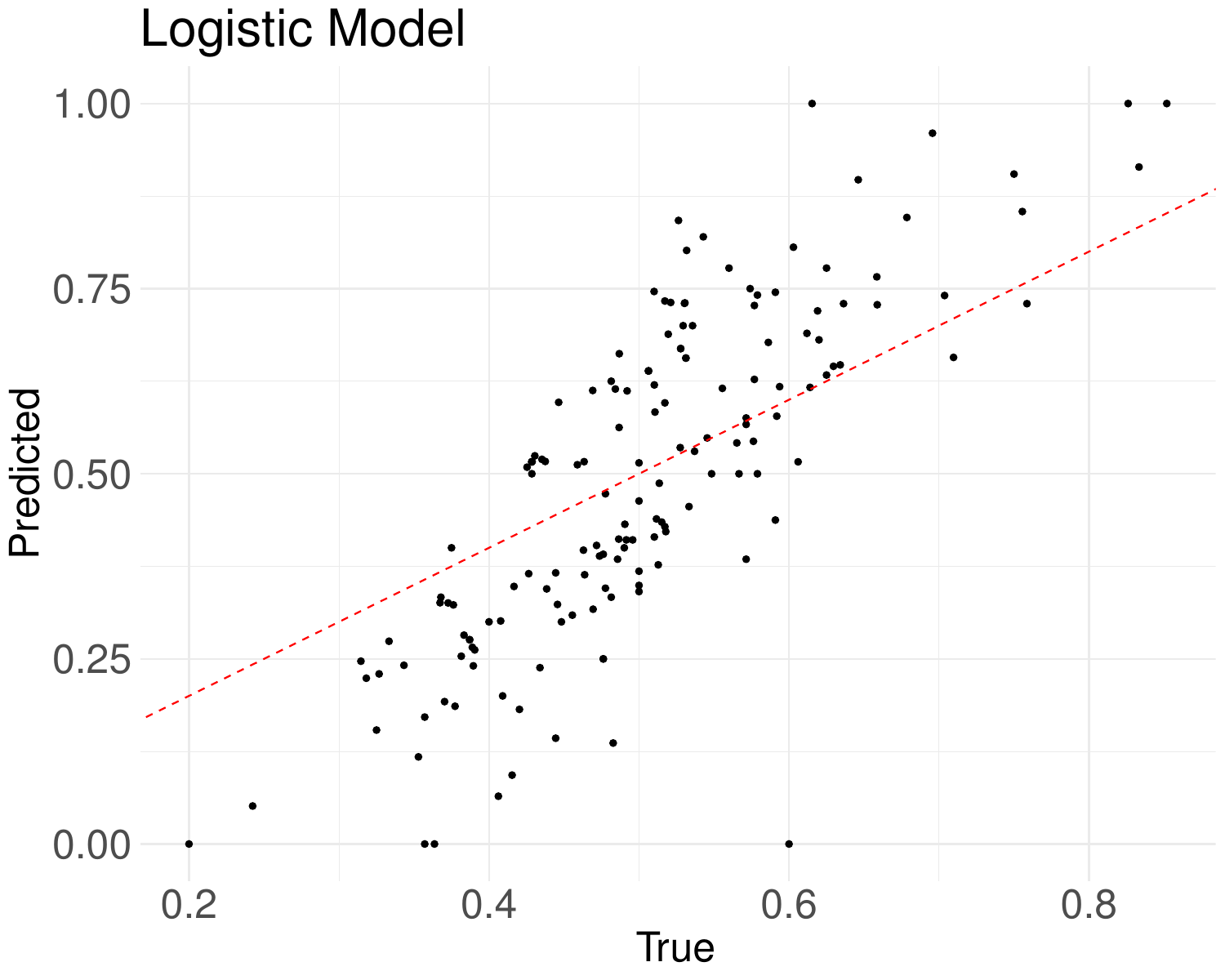}
        \label{fig:logtrain}
    }
    \vfill
    \subfigure[]{
        \includegraphics[width=0.22\linewidth,height=0.15\textheight]{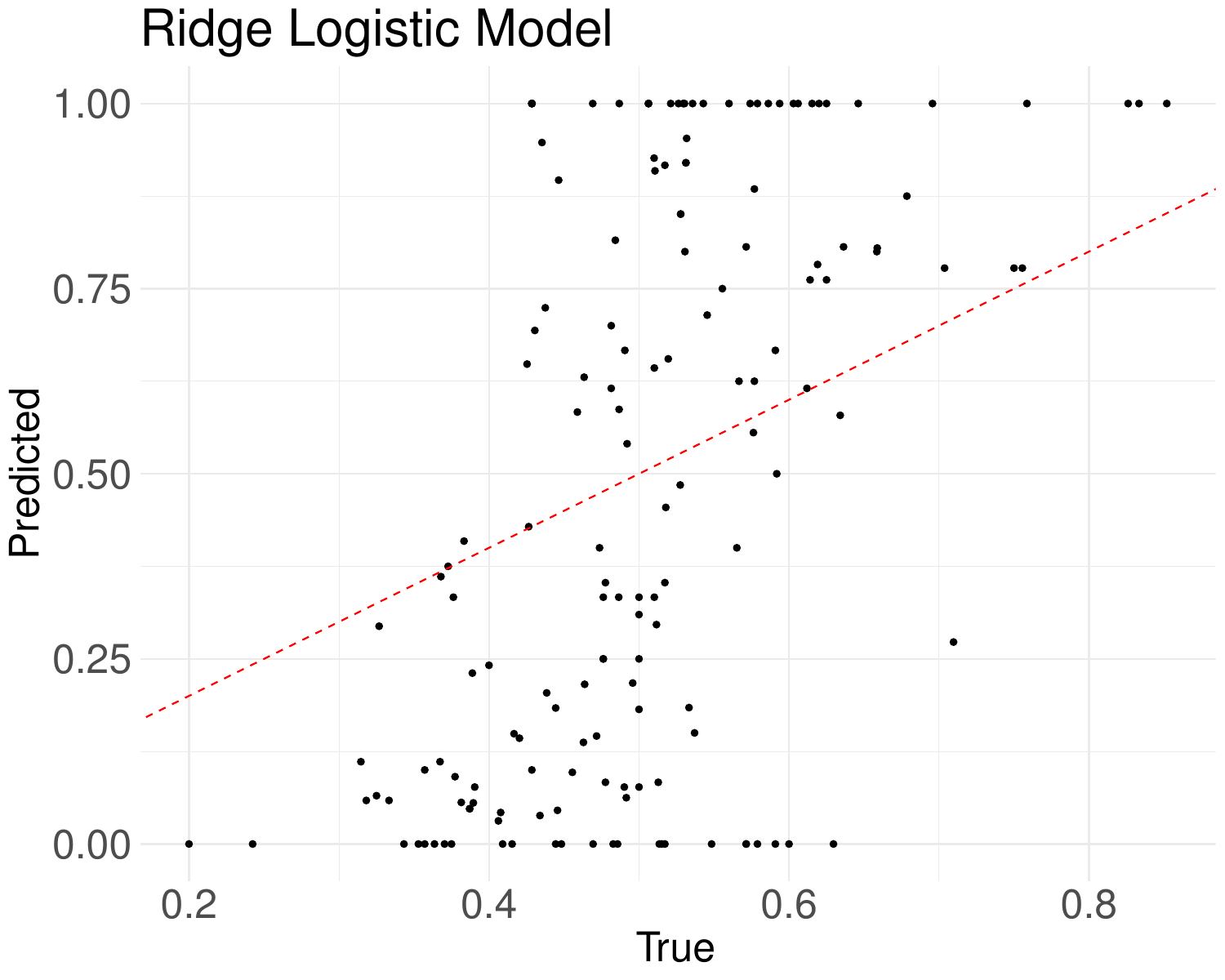}
        \label{fig:ridlogtrain}
    }
    \hfill
    \subfigure[]{
        \includegraphics[width=0.22\linewidth,height=0.15\textheight]{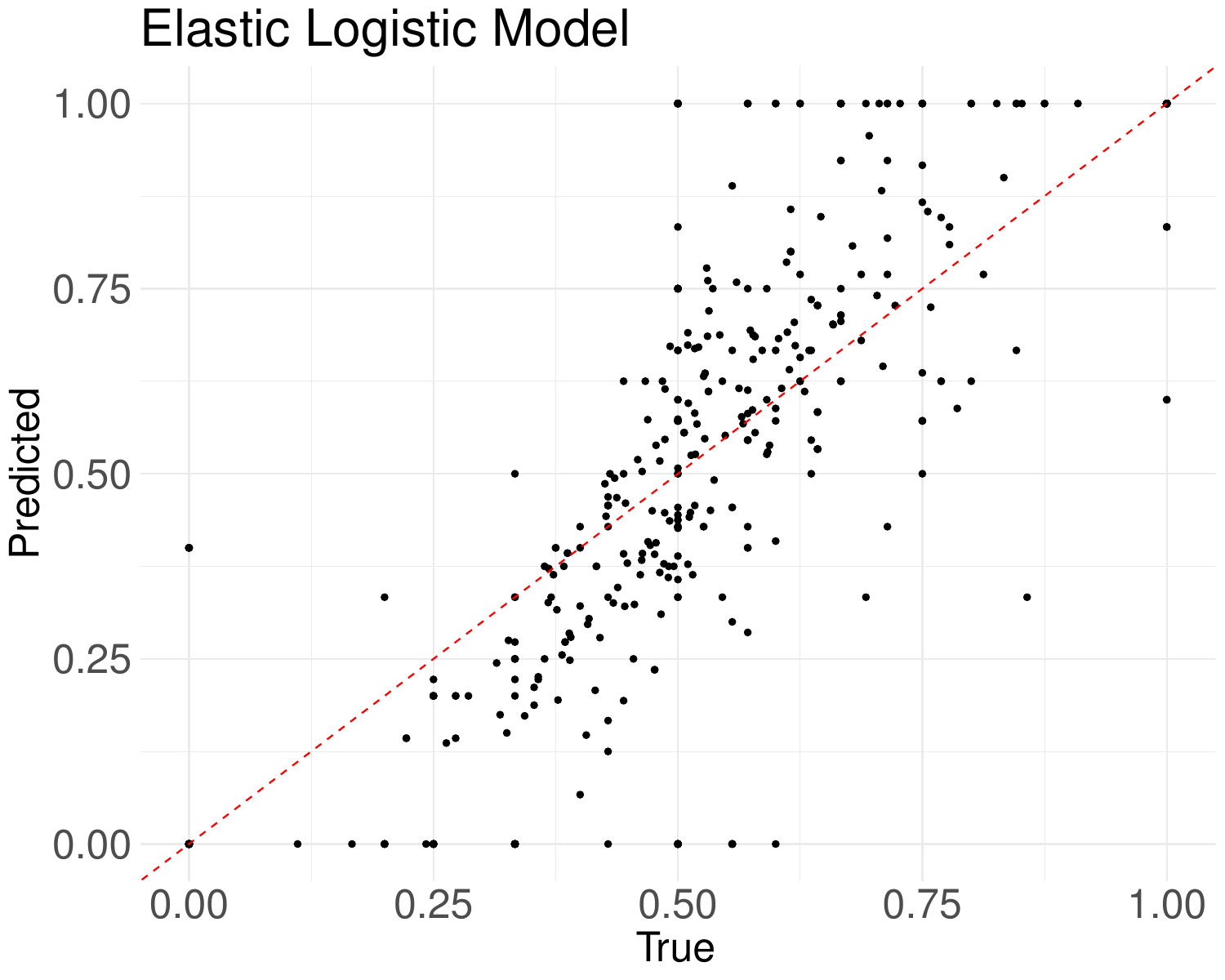}
        \label{fig:elalogtrain}
    }
    \hfill
    \subfigure[]{
        \includegraphics[width=0.22\linewidth,height=0.15\textheight]{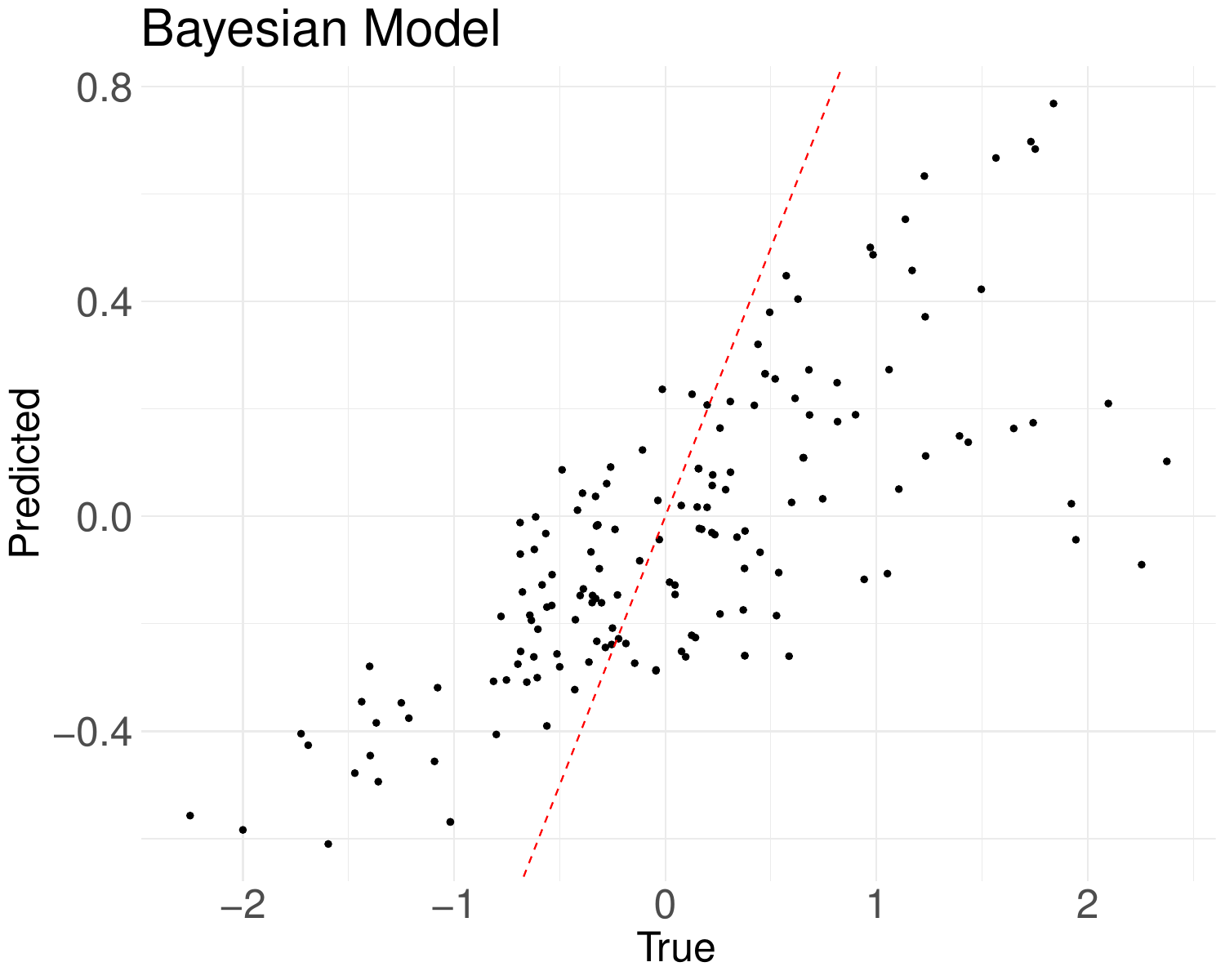}
        \label{fig:bayestrain}
    }
    \hfill    
    \begin{minipage}[b]{0.22\textwidth}
        \hfill
    \end{minipage}
    \caption{Comparison of Estimated/True Plus/Minus value (training data)}
    \label{fig:comtrain}
\end{figure*}

\begin{figure*}[!htbp]
    \centering
    \begin{minipage}[b]{0.22\textwidth}
        \subfigure[]{
            \includegraphics[width=\textwidth,height=0.15\textheight]{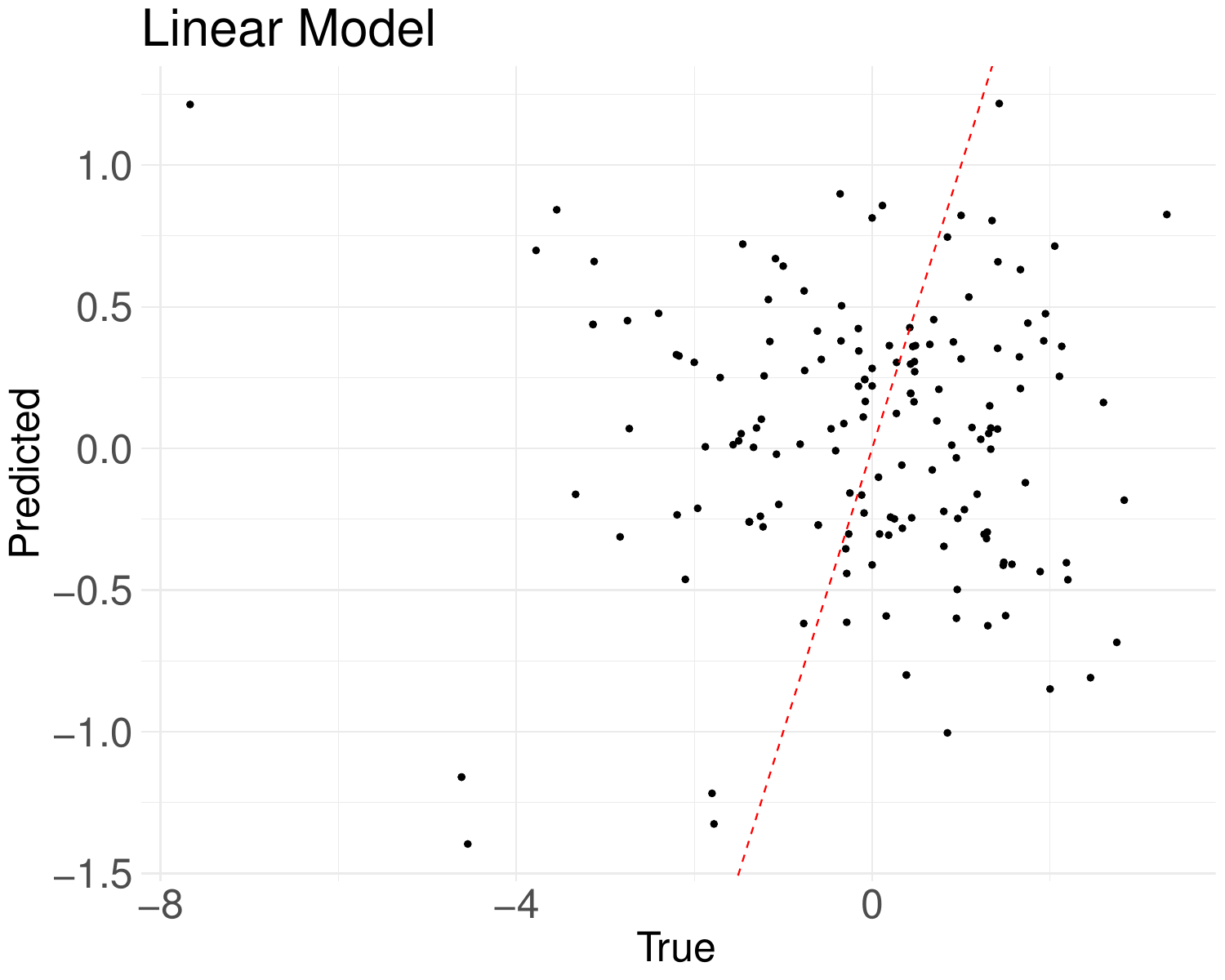}
            \label{fig:lineartest}
        }
    \end{minipage}
    \hfill
    \begin{minipage}[b]{0.22\textwidth}
        \subfigure[]{
            \includegraphics[width=\textwidth,height=0.15\textheight]{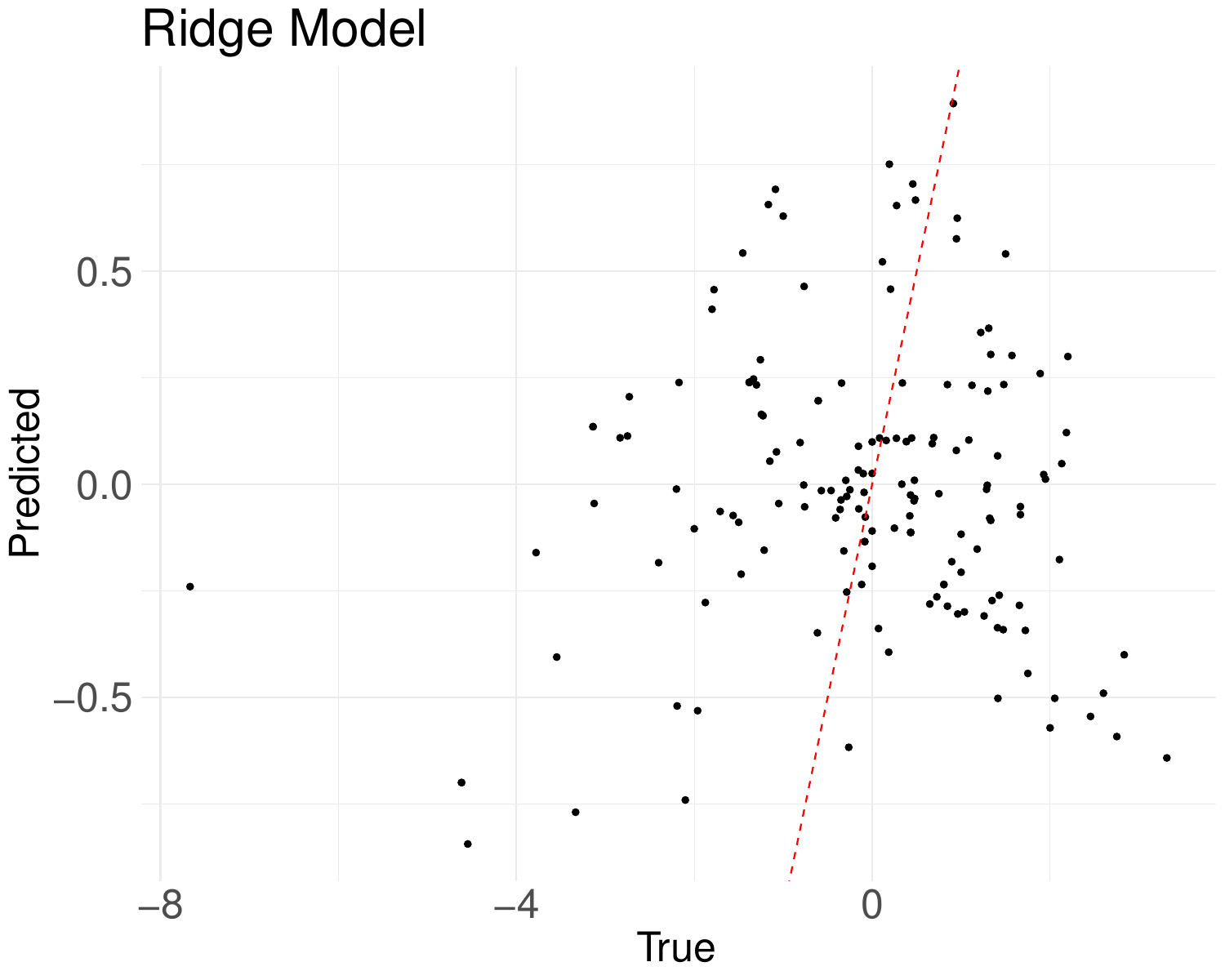}
            \label{fig:ridgetest}
        }
    \end{minipage}
    \hfill
    \begin{minipage}[b]{0.22\textwidth}
        \subfigure[]{
            \includegraphics[width=\textwidth,height=0.15\textheight]{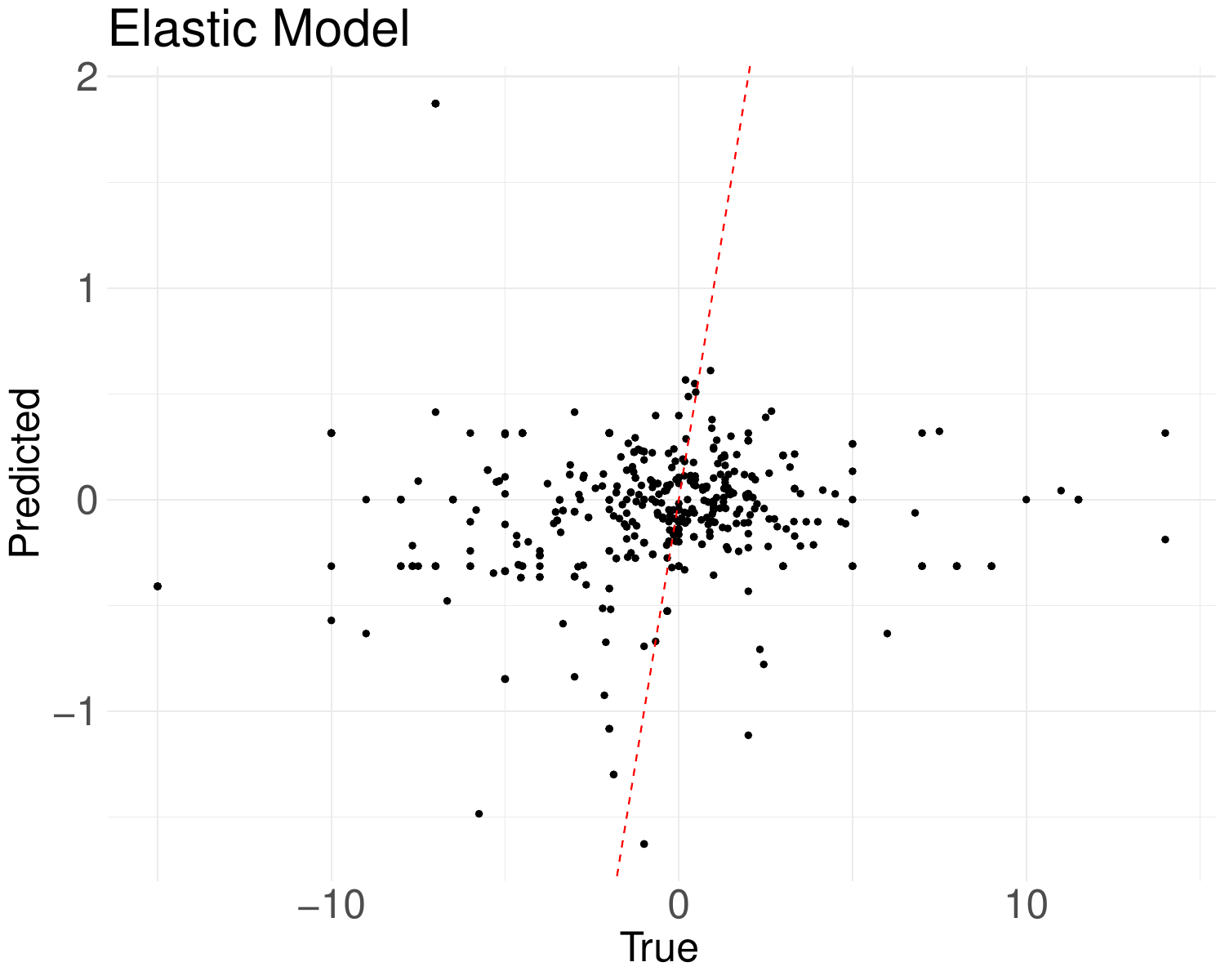}
            \label{fig:elatest}
        }
    \end{minipage}
    \hfill
    \begin{minipage}[b]{0.22\textwidth}
        \subfigure[]{
            \includegraphics[width=\textwidth,height=0.15\textheight]{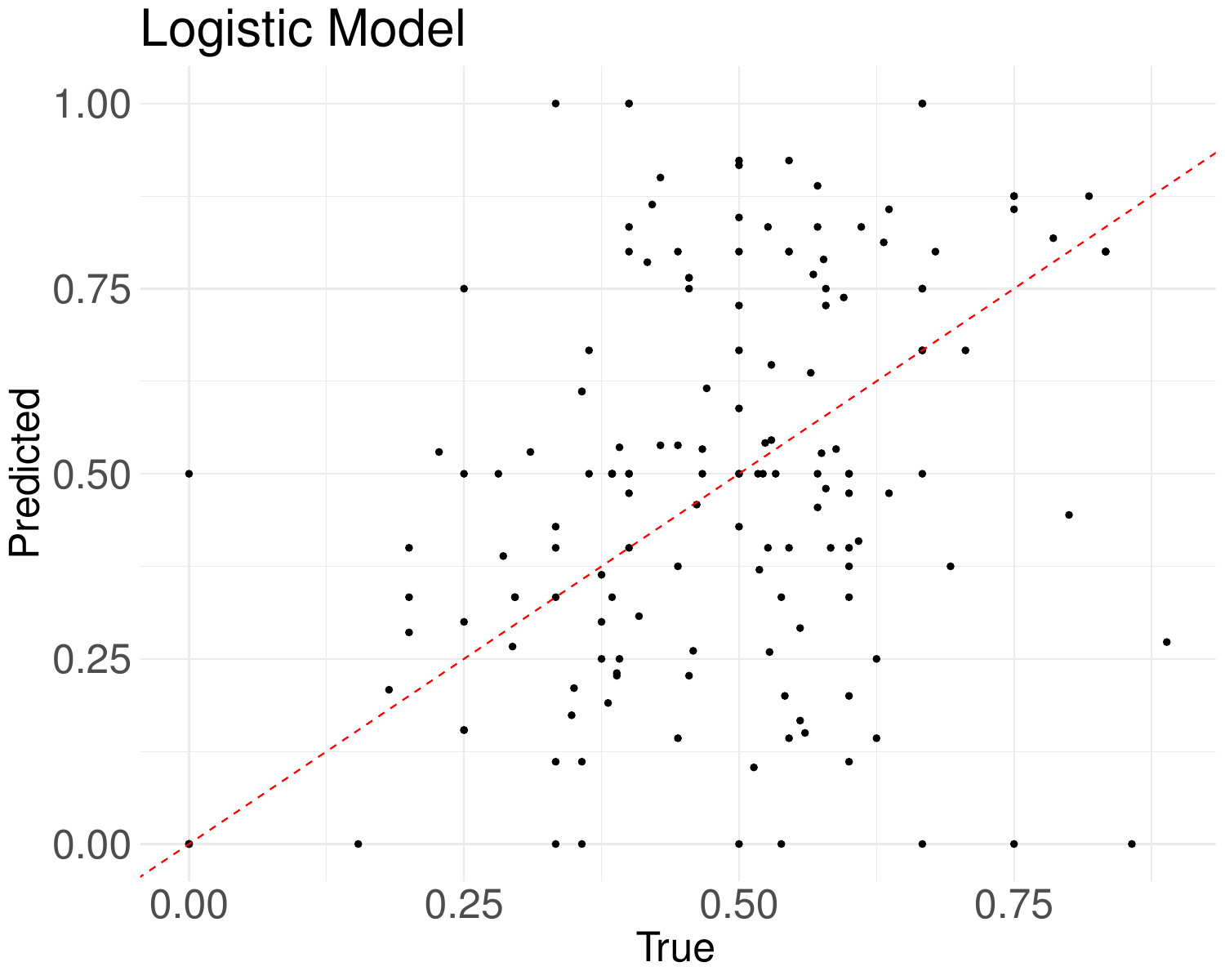}
            \label{fig:logtest}
        }
    \end{minipage}
    
    \vfill
    
    \begin{minipage}[b]{0.22\textwidth}
        \subfigure[]{
            \includegraphics[width=\textwidth,height=0.15\textheight]{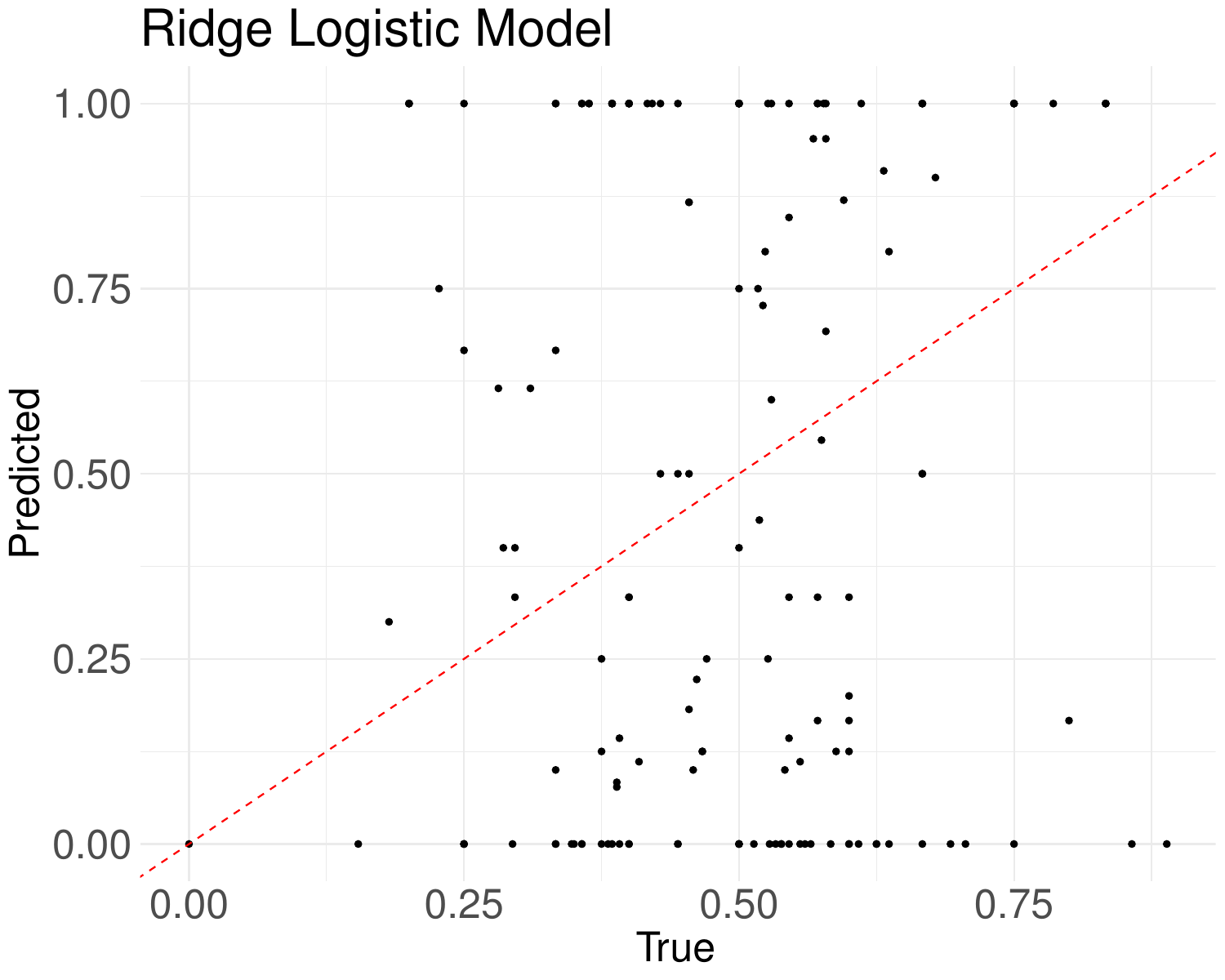}
            \label{fig:ridlogtest}
        }
    \end{minipage}
    \hfill
    \begin{minipage}[b]{0.22\textwidth}
        \subfigure[]{
            \includegraphics[width=\textwidth,height=0.15\textheight]{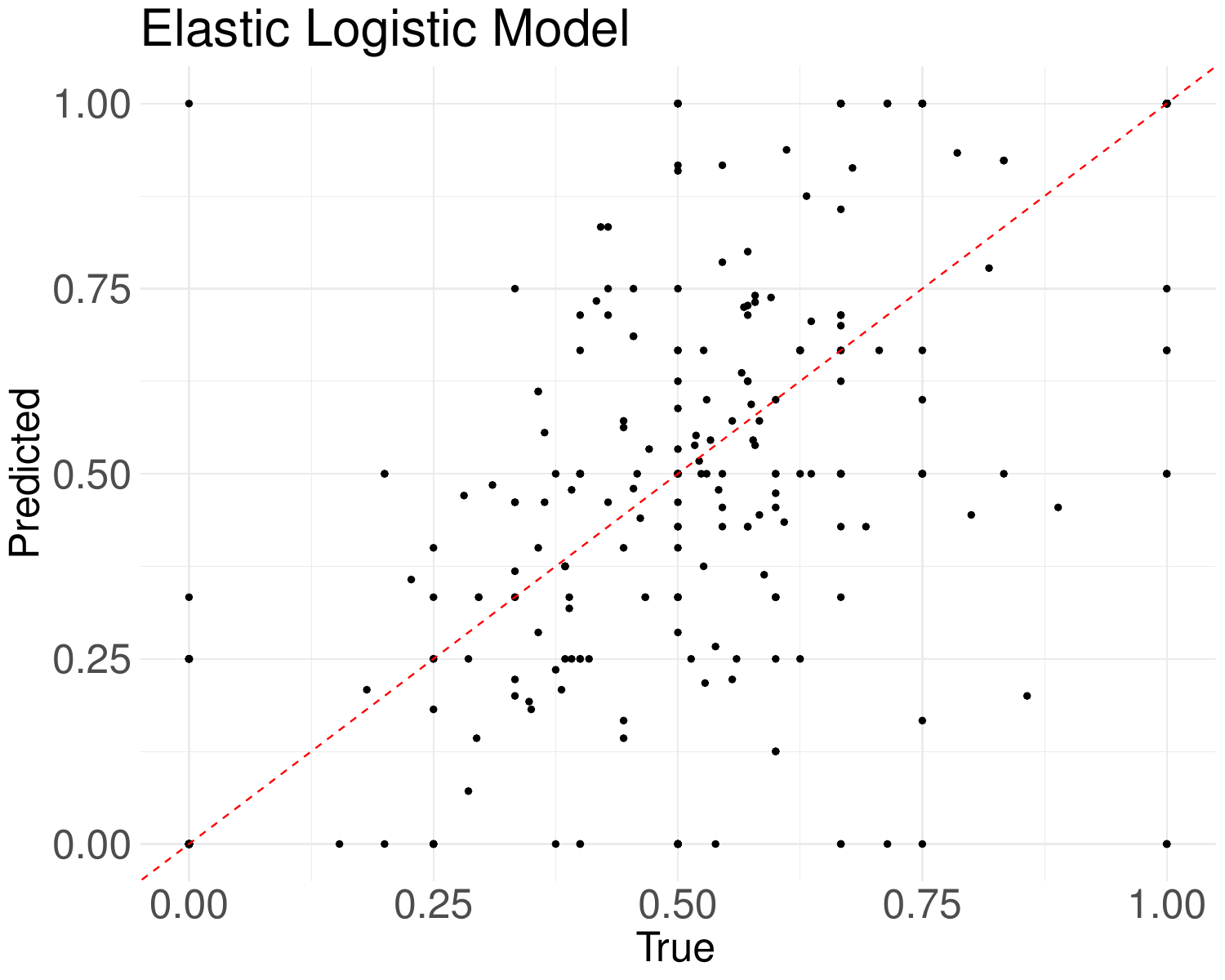}
            \label{fig:elalogtest}
        }
    \end{minipage}
    \hfill
    \begin{minipage}[b]{0.22\textwidth}
        \subfigure[]{
            \includegraphics[width=\textwidth,height=0.15\textheight]{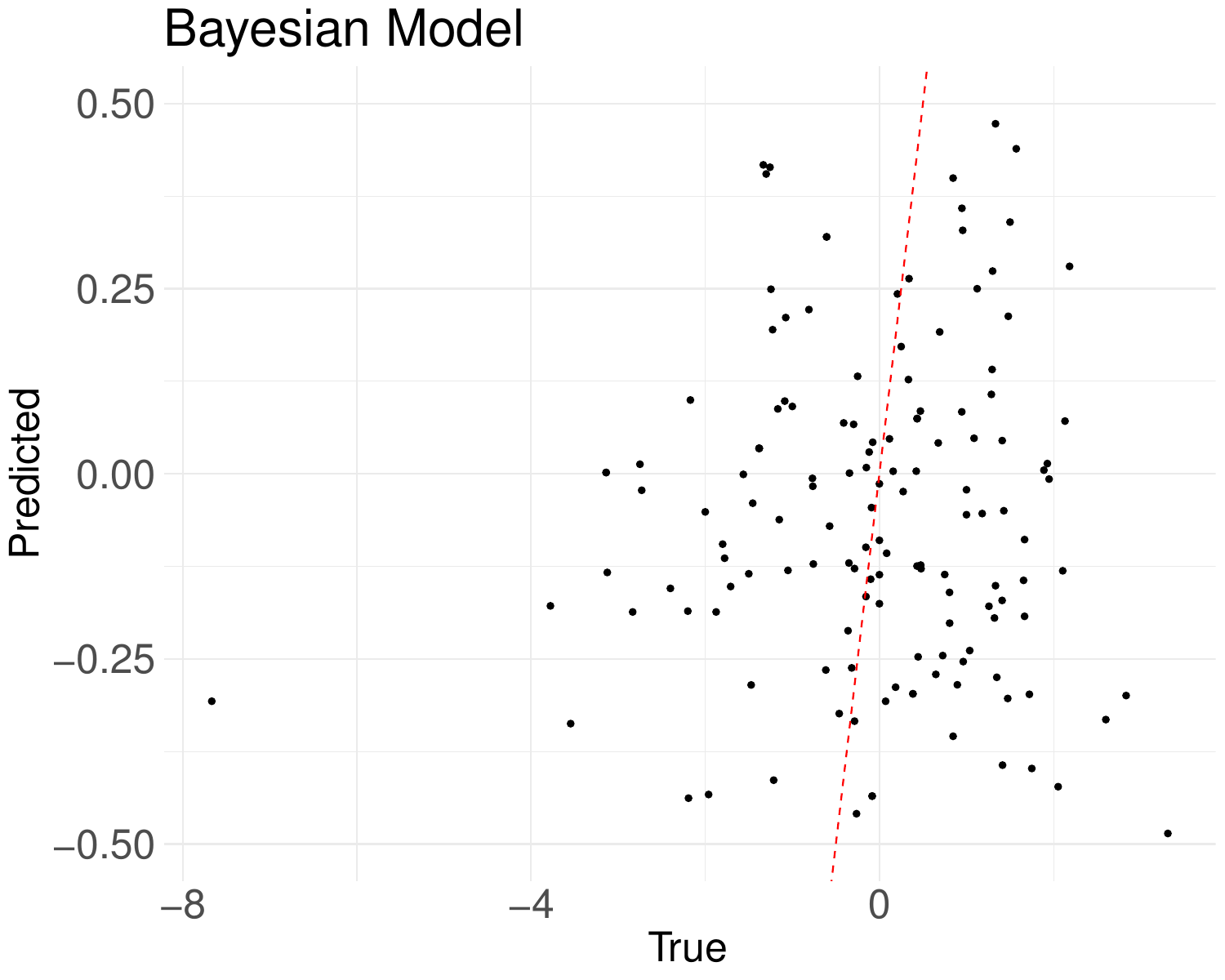}
            \label{fig:bayestest}
        }
    \end{minipage}
    \hfill
    \begin{minipage}[b]{0.22\textwidth}
        \hfill
    \end{minipage}
    
    \caption{Comparison of Predicted/True Plus/Minus value (test data)}
    \label{fig:comtest}
\end{figure*}

One major issue with all three linear regression methods is overfitting as illustrated in Figures \ref{fig:lineartrain}, \ref{fig:ridgetrain} and \ref{fig:elatrain}, where the estimated Plus/Minus values fit training data relatively well compared to the logistic and Bayesian methods. As a result, these linear regression models perform poorly on the test data as shown in Figures~\ref{fig:lineartest}, \ref{fig:ridgetest} and  \ref{fig:elatest}.
The ridge logistic model does not perform well both on training data and test data; see Figures \ref{fig:ridlogtrain} and \ref{fig:ridlogtest}.

\begin{table}[htbp]
\caption{The p-value of Pearson test to the correlation between Players' true Plus/Minus value and predicted Plus/Minus value (on test data).}
\begin{center}
\begin{tabular}{|c|c|}
\hline
\textbf{Ratings} & \textbf{p-value} \\
\hline
Ridge Plus/Minus value & 0.57 \\
\hline
Bayesian Plus/Minus value & 0.03323 \\
\hline
Logistic Plus/Minus value & 2.092e-05 \\
\hline
Elastic logistic Plus/Minus value & 2.2e-16 \\
\hline
\end{tabular}
\label{tab:pearsontab}
\end{center}
\end{table}

For remaining three models, namely, logistic, elastic logistic and Bayesian models, the p-value of Pearson test to the correlation between the players' true Plus/Minus values and the predicted Plus/Minus values on the test data are provided in Table \ref{tab:pearsontab}. As we can see, from the p-values, the predictions of these models are highly correlated to the actual Plus/Minus values. 
This suggests that players who have high ratings obtained from these three models also tend to perform well in the test dataset. This is a strong evidence to show the effectiveness of these models in predicting the performance of the players. 
In Tables~\ref{table:bapm} and \ref{table:elapm}, we further compare the elastic net logistic model and Bayesian Model.
Note that we ignored the logistic model for these comparisons, because logistic model is a special case of the elastic regression and the latter performs better than the former. 

\begin{table}[htbp]
\caption{Top 10 Players by Plus/Minus value with Their Bayesian Plus/Minus value.}
\centering
\begin{tabular}{|c|c|c|c|c|}
\hline
\textbf{Player} & \textbf{Bayesian} & \textbf{Plus/Minus} & \textbf{Bayesian} & \textbf{Rating2.0}\\
\textbf{ID} & \textbf{Rating} & \textbf{Rank} & \textbf{Rank} & \textbf{Rank} \\
\hline
xantares & 0.217 & 1 & 10 & 24 \\
\hline
interz & 0.0452 & 2 & 59 & 106 \\
\hline
n0rb3r7 & 0.0183 & 3 & 69 & 131 \\
\hline
ax1Le & 0.323 & 4 & 3 & 8 \\
\hline
fins & 0.0763 & 5 & 39 & 89 \\
\hline
nafany & 0.0964 & 6 & 31 & 95 \\
\hline
s1co & -0.0144 & 7 & 92 & 146 \\
\hline
sh1ro & 0.35 & 8 & 2 & 3 \\
\hline
m0NESY & 0.289 & 9 & 4 & 6 \\
\hline
dexter & 0.0582 & 10 & 50 & 151 \\
\hline
\end{tabular}
\label{table:bapm}
\end{table}

\begin{table}[htbp]
\caption{Top 10 Players by Plus/Minus value with Their Elastic Logistic Plus/Minus value Scores.}
\centering
\begin{tabular}{|c|c|c|c|c|}
\hline
\textbf{Player} & \textbf{Ela Logistic} & \textbf{Plus/Minus} & \textbf{Ela Logistic} & \textbf{Rating2.0}\\
\textbf{ID} & \textbf{Rating} & \textbf{Rank} & \textbf{Rank} & \textbf{Rank} \\
\hline
xantares & 0.886 & 1 & 7 & 24 \\
\hline
interz & 0.0808 & 2 & 43 & 106 \\
\hline
n0rb3r7 & - & 3 & 72 & 131 \\
\hline
ax1Le & 0.562 & 4 & 14 & 8 \\
\hline
fins & 0.982 & 5 & 4 & 89 \\
\hline
nafany & -0.171 & 6 & 117 & 95 \\
\hline
s1co & - & 7 & 100 & 146 \\
\hline
sh1ro & - & 8 & 73 & 3 \\
\hline
m0NESY & 0.919 & 9 & 5 & 6 \\
\hline
dexter & 0.0665 & 10 & 48 & 151 \\
\hline
\end{tabular}
\label{table:elapm}
\end{table}


We calculate players' Plus/Minus value, coefficient in Bayesian model as Bayesian rating and Rating2.0 in the whole data set. 
Specifically, Table~\ref{table:bapm} compares the Bayesian ranking and Rating2.0 rank for the top 10 players with the highest Plus/Minus ranking. 
We observe from this table that the Bayesian ranks are favoring players with both high Rating2.0 and high Plus/Minus values. Similarly, Table~\ref{table:elapm} compares the elastic net logistic ranks and Rating2.0 ranks of the same 10 players.
Lack of elastic net logistic ratings for some players (marked with - in the second column) indicates that those players are excluded due to the $\mathcal{L}_1$ regularization. 

In conclusion, our analysis indicates that ratings generated from the Bayesian and the elastic net logistic models are comparatively accurate overall.

\section{Conclusion}
\label{sec:conclusion}
CS:GO, as one of the most popular e-sport games in the world, seems to have a lack of fair player evaluation mechanism despite having a massive market size, and this causes difficulties for many e-sport clubs in managing their operations. This study aims to build new rating mechanisms that can analyze the impact of CS:GO players on team victories. These ratings can provide insights in a manner that can assist clubs in making informed decisions regarding player recruitment and salary evaluation to enhance their operating performance. Additionally, it offers a new perspective for selecting player awards, which could contribute to the growth of the e-sport industry. Various statistical models including regularized linear models, logistic regression models and Bayesian linear models are utilized to investigate how a player’s presence influences score and hence winning of his team. Ultimately, the elastic logistic regression model and the Bayesian linear regression model are chosen for assessing player ratings. Compared to Rating2.0, which currently the most popular and widely-used player rating system for CS:GO, our ratings have strong correlations with their teams' point differences. Further, the proposed ratings are shown to have robust abilities for predicting players' Plus/Minus values. We are confident that these new approaches provide a better evaluation mechanism for CS:GO players’ contributions to team victories than the existing methods.

\subsection{Future Research}
In the Bayesian model, we used Rating2.0 as prior for the linear regression coefficients. As a box score based data, Rating2.0 may not be the best choice. 
We believe that a future research considering a prior based on more comprehensive and detailed information about each player's position in match, duration of the play, etc. could enhance the Bayesian model.


Another important future research direction can focus on refining the current box score rating system Rating2.0. In basketball research, analysts have utilized Rosenbaums APM model \cite{APM} on player ratings to identify which box scores better fit players’ impact on the game results, and built a new box score based metric, called {\em Box Plus/Minus value} \cite{bpm}, to provide real-time player ratings for basketball matches. However, as far as we are aware, there is no such research conducted on e-sports. 
A future study can aim to refine Rating2.0 using the Plus/Minus values from proposed elastic logistic regression and the Bayesian regression models. This could be expected to make better evaluations on players’ contribution to team victories of the matches played over a short period of time, and thus eventually helping in development of e-sport industry.


\bibliographystyle{IEEEtran}
\bibliography{ref}

\end{document}